\documentclass[final,longbibliography,12pt]{revtex4-1}

\usepackage{graphicx}
\usepackage{amssymb}
\usepackage{amsfonts}
\usepackage{amsmath}
\usepackage{slashed}

\newcommand{\be}{\begin{equation}}
\newcommand{\ee}{\end{equation}}
\newcommand{\wf}{wavefunction\;}
\newcommand{\wfs}{wavefunctions\;}
\newcommand{\infl}{influence function\;}

\newcommand{\rmi}{\mathrm{i}}

\newcommand{\dg}{\dagger}
\newcommand{\fy}{\slashed}

\begin{document}

\title{Spin-Statistics Connection for Relativistic Quantum Mechanics}


\author{A. F. Bennett}

\email{bennetan@oregonstate.edu}

\affiliation{College of Earth, Ocean and Atmospheric Sciences\\Oregon State University\\104 CEOAS Administration Building\\ Corvallis, OR 97331-5503, USA}



\date{\today}


\begin{abstract}
The spin-statistics connection has been proved for nonrelativistic quantum mechanics (Jabs, A., 2010: Found. Phys, {\bf 40}, 776-792). The proof is extended here to the relativistic regime using the parametrized Dirac equation. A causality condition is not required.


\end{abstract}

\maketitle

\section{Introduction}\label{sec:Intro}
The spin-statistics connection  is regarded as one of the most important results in theoretical physics   \cite{IZ,Wei95,Zee03,Sr07}. The standard proof  in Quantum Field Theory  requires relativistic physics, yet it has been argued that spin is intrinsically a nonrelativisitic phenomenon \cite{WeiQM13} since  it characterizes the representations of $SO(3)$\,.  On the other hand the electron gyromagnetic ratio $g=2$ is a consequence of the relativistic wave equation of Dirac, but must be specified in the nonrelativistic wave equation of Pauli \cite{BjDr64}.

There is an elementary proof in nonrelativistic quantum mechanics for the spin-statistics connection \cite{Jabs10}. The objective of this investigation is an elementary extension of the proof  to the relativistic regime, using the parametrized Dirac equation attributed to Feynman and others \cite{Alv98}. Jabs \cite{Jabs10} briefly outlines an alternative relativistic approach involving helicity, but the helicity operator is only Lorentz-invariant for massless particles. The argument here follows naturally from the nonrelativistic proof, and involves the relativistic spin operator which is Lorentz-invariant regardless of mass.

The contents of this article are as follows. Section 2 includes a statement of the single-particle parametrized Dirac equation with an outline of its utility, the forms of the free plane wave solutions, the splitting of positive-energy plane waves into positive-energy waves propagating forward and backward in coordinate time, and the equation for many particles. The eigenstates of the spin operator are defined in Section 3, where it is proved that any free state at any parameter value evolves from a uniquely determined free spin eigenstate prepared at an earlier (algebraically smaller) parameter value. The proof of the nonrelativistic spin-statistics connection by Jabs \cite{Jabs10}  is extended in Section 4 to the relativistic regime at the preparation parameter value. The multiple-particle analysis of Jabs is repeated, for the case of two spin-1/2 particles prepared in spin eigenstates, without a significant loss of generality but with a significant gain in simplicity. The free two-particle influence function for other parameter values, while mixing the spin eigenstates for each particle, preserves the antisymmetric or fermionic form of the two-particle state. The result is discussed in the concluding Section 5, in particular the lack of need for a causality condition.

\section{The Parametrized Dirac Equation}

\subsection{Covariant Formulation}\label{S:covform}

 The \wf for a single spin-1/2 particle  is a four-spinor  $\psi(x,\tau)$. The event $x$ is in ${\mathbb R}^4$,  the parameter  $\tau$ is an independent variable in ${\mathbb R}$.  The event $x$ is also denoted by $x^\mu$ having indices $\mu=0,1,2,3$, with $x^0=ct$ where $c$ is the speed of light and $t$ is coordinate time. The Lorentz metric $g^{\mu\nu}$ on  ${\mathbb R}^4$ has  signature $(- + + +)$. The position $\bf{x}$  is denoted by $x^j$ having indices $j=1,2,3$\,, and so  $x=(ct,\bf{x})$.

The parametrized Dirac equation for $\psi$ is  
\be\label{PDIR}
\frac{\hbar}{\rmi c}\frac{\partial}{\partial \tau}\psi+\gamma^\mu\Big(\frac{\hbar}{\rmi}\frac{\partial}{\partial x^\mu}-\frac{e}{c}A_\mu\Big)\psi=0
\ee
where $\hbar$ is the reduced Planck's constant,  and $e$ is the charge of the particle, while the $\gamma^\mu$ are the four Dirac matrices. The  Maxwell electromagnetic potential $A^\mu(x)$ is  independent of the parameter $\tau$. The parametric role of $\tau$ is clear for a wave packet:  at large  negative $\tau$ the packet may be localized in one region of space-time, and at large positive $\tau$ in another region. That is,  $\tau$ parametrizes the evolution of the space-time moments of the packet.  The covariance of the theory with respect to the homogeneous Lorentz transformation $(x^\mu)' =\Omega^{\mu}_{\;\;\nu}x^\nu$\,, represented by $ \psi'(x,\tau)=S(\Omega)\psi(\Omega^{-1}x,\tau)$\,, follows for $S(\Omega)$  generated in the standard way \cite{BjDr64}. There is no mass constant  in (\ref{PDIR}), but masses are introduced through boundary conditions as $\tau\to \pm \infty$\,. Feynman's development of Quantum Electrodynamics using (\ref{PDIR}) is reviewed by Garcia Alvarez  and Gaioli \cite{Alv98}. A simple consequence of (\ref{PDIR}) is the identity
\be\label{VCUR}
\frac{\partial}{\partial \tau}\overline{\psi}\psi+\frac{\partial}{\partial x^\mu}j^\mu=0
\ee
where $\overline{\psi} = \psi^\dg\gamma^0$, and the $\tau$-dependent current is $j^\mu= c\overline{\psi}\gamma^\mu \psi$. The indefiniteness of the conserved, invariant bilinear form $\overline{\psi}\psi$ has impeded  the development of the parametrized Dirac formalism as a relativistic extension of quantum mechanics \cite{Barut85,Evans98}. In the case of a  \wf having `sharp mass $m_e$' as in $\psi(x,\tau)=\exp(\rmi m_e \tau)\psi(x,0)$\,, and low kinetic energy $|{\mathbf p}|^2 \ll m_e^{\,2}$ where ${\mathbf p}$ is a three-momentum representative of the wavefunction, the Dirac four-spinor $\exp(\rmi m_e x^0)\psi(x,0)$ may be projected onto  a `large upper' and a `small lower' Pauli two-spinor, both varying slowly in coordinate time $x^0$. The `large upper' component, denoted $\varphi$ in \cite{BjDr64}, satisfies the Pauli or spin-1/2 Schr\"{o}dinger equation   \cite[Ch. 1]{BjDr64} which  conserves the Hermitian form $\varphi^\dg\varphi$\,. The situation here is as for quantum mechanics (QM) and Quantum Field Theory (QFT): at low energy the parametrized Dirac equation has a Born interpretation, but at high energy it yields only transition amplitudes. 

The discrete symmetries for (\ref{PDIR}) are given in \cite{Benn2014}\footnote{For minor corrections to that article see arXiv:1406.0750}. The electromagnetic potential $A^\mu$ is calculated semiclassically in \cite{Benn2014}, that is, the the source for the potential is a M\o ller current \cite{BjDr64,Tay72}. An expansion of (\ref{PDIR}) in powers of the fine structure constant $\alpha=e^2/4\pi \hbar c$ yields the Mott scattering cross-section at first order. Combined with a partial summation that is justified for weak scattering, expansion to higher orders yields the Uehling potential, the self-energy and self-mass  of fermion lines, the anomalous magnetic moment of the electron, the Lamb shift  and the axial anomaly of Quantum Electrodynamics (QED) \cite{Benn2014}. The standard cross sections for pair interactions, such as electron-positron annihilation, may be  derived from the multiple-particle parametrized Dirac equation (see Section \ref{S:Many}) without recourse to hole theory. In brief, two particles are interacting. The real  electron scatters off a real photon into a virtual electron, and similarly for the positron. Summing the products of the two amplitudes over the virtual energy-momenta yields the inertial fermion line in the standard diagram for pair annihilation into two photons. The multiple-particle parametrized Dirac equation also yields the Bethe-Salpeter equation for bound states without further conjecture \cite{Benn2014}. Semiclassical electrodynamics in the presence of an isotropic  zero-point electromagnetic far field having energy density $\hbar\omega/2$ at frequency $\omega$ yields the blackbody spectrum, the van der Waals forces, the Casimir effect, the Einstein `$A$' coefficient for spontaneous emission and the photoelectric effect \cite{Boy69,Boy75,Camp99,Fox06,GreZaj05}. It is widely held that a classical theory of radiation cannot explain the sub-Poisson statistics and antibunching routinely observed \cite{Thorn04} in photon counting at very low levels of illumination, but see \cite[Ch. 13]{Pena96}. In summary, the parametrized Dirac equation for spin-1/2 particles interacting through a semiclassical potential has extensive utility. 

The units are now chosen such that  $c=\hbar=1$. The summation convention is assumed for  Greek indices such as $\mu =0,1,2,3$\,. The covariant and contravariant indices $\mu,\nu,\dots$ will be omitted where convenient, as in $x= x^\mu$\,, $p = p^\mu$\,, $p\cdot x = p^\mu x_\mu$\,.

\subsection{Plane Waves}\label{S:plwvs}
Plane wave solutions of (\ref{PDIR}) for a vanishing potential $A^\mu
=0$ have the forms
\be\label{planes}
\psi_\pm^{(r)}(x,\tau)=w_\pm^{(r)}(p) \exp[\rmi(p\cdot x \pm \varphi_pm_p\tau)]
\ee
where $w_\pm^{(r)}(p)$  are covariant Dirac four-spinors for $r=1,2$\,,  $m_p=\sqrt(-p\cdot p) >0$ is the positive mass for a subluminal state,  and $\varphi_p={\rm sign}(p^0)=p^0/E_p$ where $p^0$ is the energy with  magnitude $E_p=|p^0|$ .  The phases of the complex exponentials are  $p\cdot x \pm \varphi_pm_p\tau$, hence the rates of change of the coordinate time $t=x^0$ with respect to the parameter $\tau$ at constant phase and position ${\mathbf x}$ are
\be\label{fwdbwd}
\frac{dt}{d\tau}=\pm m_p/E_p
\ee
regardless of the value of $\varphi_p$. That is, $\psi^{(+)}$ and $\psi^{(-)}$ propagate forward and backward in time respectively, regardless of the sign of the energy $p^0$. The four-spinors $w^{(r)}_{\pm}(p)$ are eigenvectors of the energy-momentum projections $\Lambda_{\pm}(p) =(1/2m_p)(m_p\mp\varphi_p\fy{p})$ which satisfy $\Lambda_++\Lambda_-=1$\,, $\Lambda_{\pm}\Lambda_{\pm}=\Lambda_{\pm}$ and $\Lambda_{\pm}\Lambda_{\mp}=0$\,. That is, $w_+^{(r)}=\Lambda_+u^{(r)}$ and $w_-^{(r)}=\Lambda_-v^{(r)}$ for some four-spinors $u^{(r)}$ and $v^{(r)}$, where $r=1,2$\,. 

\subsection{Splitting}\label{S:split}

Suppose that at $\tau=0$ the state is of positive energy and has the plane \wf \be\label{psi}
\psi(x,0)=w\exp[\rmi p\cdot x]\,,
\ee
where $p^0>0$\,, $w=({\sf w}_+{\mathbf a}_+ + {\sf w}_-{\mathbf a}_-)$,  ${\sf w}_\pm=(w_\pm^{(1)},  w_\pm^{(2)})$\,, and the coefficients ${\mathbf a}_\pm$ are any $2 \times 1$ complex matrices. Evolving the state with the free \infl $\Gamma^0_+$ as constructed in \cite{Benn2014} yields, for $\tau >0$, 
\be\label{plus}
\begin{aligned}
\psi(x,\tau)&=\frac{1}{\rmi}\int d^4x' \,\Gamma_+^0(x-x',\tau)\psi(x',0)\\
&={\sf w}_+{\mathbf a}_+ \exp[\rmi (p\cdot x+\varphi_p m_p\tau)]\,.
\end{aligned}
\ee
For $\tau>0$ the free \infl $\Gamma^0_+$ projects the prepared positive-energy \wf $\psi(x,0)$  onto the free positive-energy \wfs that propagate forward in time as $\tau$ increases. If the prepared state has negative energy then $\Gamma^0_+$ projects onto the free negative-energy \wfs that propagate backward in time. The $\mathcal{TPC}$-conjugate of a \wf for a particle is by definition the \wf for the oppositely-charged particle or `antiparticle' having energy of opposite sign and propagating in the opposite sense in time. There are antiparticles of positive and negative energy, propagating in either sense in time   \cite{Benn2014}.

\subsection{Many Particles}\label{S:Many}
The parametrized Dirac equation (\ref{PDIR}) has a natural extension for many particles in the  space spanned by tensor products of single-particle wavefunctions. A two-particle \wf $\Psi(x,y,\tau)$ satisfies \cite{Benn2014}
\be\label{TWO}
\frac{1}{\rmi}\frac{\partial}{\partial \tau}\Psi+\fy{\pi}(x)\otimes I_4\,\Psi+I_4\otimes \fy{\pi}(y)\,\Psi=0
\ee
where $\pi^\mu(x)=(1/\rmi)\partial /\partial x_\mu-e_1A^\mu(x)$ and $\pi^\mu(y)
=(1/\rmi)\partial /\partial y_\mu-e_2A^\mu(y)$\,. The free \infl $\Gamma_{++}^0(x-x',y-y',\tau-\tau')$ is  \cite{Benn2014}
\be\label{Gam02}
\Gamma_{++}^0(x-x',y-y',\tau-\tau')=\frac{1}{\rmi}\Gamma_+^0(x-x',\tau-\tau')\otimes\Gamma_+^0(y-y',\tau-\tau')\,.
\ee
If there is a nonvanishing external field $A^\mu$\,, and if $\Gamma_+(x,\tau;x',\tau')$ is the single-particle \infl constructed using $\Gamma^0_+(x-x',\tau-\tau')$ as in \cite{BjDr64,Benn2014}, then the \infl for (\ref{TWO}) is
\be\label{GamA2}
\Gamma_{++}(x,y,\tau;x',y',\tau')=\frac{1}{\rmi}\Gamma_+(x,\tau;x',\tau')\otimes\Gamma_+(y,\tau;y',\tau')\,.
\ee

\section{Spin Eigenstates}\label{S:spin}

For any spacelike four-vector $s^\mu$\,, with $s\cdot s=+1$, there is the spin operator $S(s)=(1/2)\fy{s}\gamma^5$ (not to be confused with $S(\Omega)$, the standard notation for the four-spinor representation of the Lorentz transformation $\Omega$).  The eigenvalues of $S(s)$ are $l=\pm1/2$\,. The associated projections are $P(s)=S(s)+1/2$ and $Q(s)=1-P(s)$\,. The spin operator $S(s)$  commutes with the energy-momentum projections $\Lambda_{\pm}(p)$ if and only if the spin axis and the energy-momentum are orthogonal. That is, $[S(s),\Lambda_{\pm}(p)]=0 \iff  p \cdot s=0$\,. Owing to the two-fold degeneracy of both eigenvalues of $S(s)$ and also of both eigenvalues of say $\Lambda_+(p)$, an eigenvector of either operator is not necessarily an eigenvector of the other even if $p\cdot s=0$\,.

Consider the case in which $p \cdot s \ne 0$\, and the \wf $\psi(x,\tau)$  defined in (\ref{plus})  for $\tau>0$ is not an eigenvector of $S(s)$\,. It is  shown in the Appendix that for any ${\mathbf a}_+$ as in (\ref{plus}) there is a unique solution for ${\mathbf a}_-$ such that the four-spinor $ w={\sf w}_+{\mathbf a}_+ + {\sf w}_-{\mathbf a}_-$  satisfies $S(s) w=+(1/2) w$\,. That is, for any positive-energy, forward-propagating plane \wf $\psi(x,\tau)$\,, where $\tau>0$\,, there is a unique, positive-energy plane \wf $\psi(x,0)$ such that  $S(s)\psi(x,0)=(1/2)\psi(x,0)$ and  $ \psi(x,0)$ evolves into $\psi(x,\tau)$ for $\tau>0$\,. 

If $p\cdot s=0$ there is in general no solution for  ${\mathbf a}_-$ such that the \wf $\psi(x,0)$ is an eigenvector of $S(s)$\,. However, given a timelike energy-momentum $p^\mu$\,, a spacelike spin axis $s^\mu$ may be chosen in  a non-rest frame such that $p\cdot s \ne 0$\,.

Finally consider, for $\tau>0$,  a positive-energy, forward-propagating, separable two-particle plane \wf  $\Psi(x,y,\tau)$\, with energy-momenta $p$ and $q$ respectively, where $p^0>0$ and $q^0>0$\,. Owing to the separability of the two-particle \infl (\ref{Gam02}) there is a positive-energy  plane \wf $\Psi(x,y,0)$ such that, for some spin axis $s^\mu$\,, $ S(s) \otimes S(s)\,\Psi(x,y,0)=(1/4)\Psi(x,y,0)$ and  $\Psi(x,y,0)$ evolves into $\Psi(x,y,\tau)$ for $\tau > 0$\,. Note especially that the same spin axis $s^\mu$  is specified for both particles. The construction may be extended to any number of particles, with only a single spin axis $s^\mu$ specified. As discussed in the Appendix, the  case $p\cdot s=0$ for any subset  of many particles can be avoided by an appropriate choice of the single spin axis $s^\mu$\,.

It need not be assumed that mass is an intrinsic property of a particle. If it is so assumed  then indistinguishable particles all have the same mass $m_e$\,, and $m_e^2=-p\cdot p$  for all plane \wfs with energy-momentum $p^\mu$\,. Free \wfs including those prepared as spin eigenvectors at $\tau=0$ then satisfy the free Klein-Gordon equation \cite{BjDr64}. It may be recalled that free solutions of the sharp-mass Dirac equation are also free solutions of the Klein-Gordon equation, but not vice-versa.

\section{Spin and Statistics}\label{S:spinstat}

\subsection{At $\tau=0$}\label{zero}

\subsubsection{same spin}\label{same}
Consider many free, indistinguishable spin-1/2 particles.  Assume further that the \wfs are all  spin eigenvectors for the same spin axis $s^\mu$\,, all with the same eigenvalue $l=-1/2$ or $l=+1/2$\,. Note that a spin axis $s^\mu$ has been specified but, as pointed out by Jabs \cite{Jabs10},  not a spin frame. Thus the referencing of each single-particle \wf to a common frame involves a spatial rotation through an angle $\chi$, which can be restricted to the range $[0,2\pi]$\,. It follows \cite{BjDr64} that the rotation augments the phase of that spin-1/2 \wf  by $l\chi$\,. So in any common spin frame each single-particle \wf is defined only to within a factor of $\exp(\rmi  l \chi)$\,. This is precisely the same as the nonrelativistic situation investigated by Jabs \cite{Jabs10}\,, who points out that indistinguishability requires the exchange of particles to involve also an exchange of frame-dependent angles. That exchange can be made by, say,  a counterclockwise rotation. Demanding further that the exchange be made in an homotopically consistent way has significant consequences.  For example, consider the separable  two-particle \wf 
\be\label{ex0}
\Psi=\psi(x,\tau, l, \chi)\otimes \phi(y,\tau, l, \lambda)\,.
\ee
Jabs \cite{Jabs10} emphasizes that the frame angles $\chi$ and $\lambda$ are parameters rather than observables. First, exchange the \wfs $\psi$ and $\phi$ but not the phases $\chi$ and $\lambda$, and then add the result to $\Psi$ to obtain 
\be\label{ex1}
\Psi_{S'}=\psi (x,\tau, l,\chi)\otimes \phi(y,\tau,l, \lambda)+\phi (x,\tau ,l, \chi)\otimes \psi(y,\tau , l,\lambda)\,.
\ee
Such simple addition is stipulated for the \wfs of indistinguishable particles, whenever a transition to either state is possible. Second, exchange the frame angles in the rightmost summand in (\ref{ex1}) by, say, counterclockwise rotation. If, say,  $\chi<\lambda$,  run $\chi$ through $\lambda-\chi$ to obtain a rotation factor $\exp[\rmi l(\lambda-\chi)]$\,, and run $\lambda$ through $2\pi -\lambda+\chi$ to obtain a factor $\exp[\rmi l(2\pi  -\lambda +\chi)]$\,. Then (\ref{ex1}) becomes 
\be\label{ex2}
\Psi_S=\psi (x,\tau, l,\chi)\otimes \phi(y,\tau, l,\lambda)-\phi (x,\tau ,l,\lambda)\otimes \psi(y,\tau ,l, \chi)\,.
\ee
The frame-angle dependence of the \wfs $\psi$ and $\phi$ are of the form $\psi(x,\tau ,l, \chi)=\psi(x,\tau,l)\exp(\rmi l\chi)$ and  $\phi(y,\tau ,l,\lambda)=\phi(y,\tau,l)\exp(\rmi l\lambda)$\,, hence referring the \wfs to their original spin frames yields the common factor $\exp[-\rmi l(\chi + \lambda)]$ for the two summands in (\ref{ex2}). The common factor and the frame-dependent angles $\chi$ and $\lambda$  may therefore be ignored in the calculation of observable quantities, such as cross-sections, using (\ref{ex2}). In conclusion, the \wf for two indistinguishable particles  in the same spin eigenstate is antiysmmetric with respect to particle exchange \cite{Jabs10}. 

\subsubsection{different spins}\label{diff}
Consider two otherwise indistinguishable particles in eigenstates of opposite spin. The separable two-particle \wf is 
\be\label{exd0}
\Psi=\psi(x,\tau,l,\chi)\otimes\phi(y,\tau,n,\lambda)
\ee
where $|l|=|n|=1/2$ and $l=-n$\,. Exchanging \wfs but not angles and then adding yields 
\be\label{exd1}
\Psi_{S'}=\psi (x,\tau, l,\chi)\otimes \phi(y,\tau,n, \lambda)+\phi (x,\tau ,n, \chi)\otimes \psi(y,\tau , l,\lambda)\,.
\ee
An homotopically-consistent  exchange of angles in the rightmost summand in (\ref{exd1})  assuming, say, counterclockwise rotation and $\chi <\lambda$ yields 
\be\label{exd2}
\Psi_{S\,}=\psi (x,\tau, l,\chi)\otimes \phi(y,\tau,n, \lambda)-\kappa\phi (x,\tau ,n,\lambda)\otimes \psi(y,\tau ,l, \chi)\,,
\ee
where $\kappa=\exp[\rmi(l-n)(\lambda-\chi)]$. The \wf $\Psi_S$ in (\ref{exd2}), which of course is valid also for the case $l=n$\,, is not antisymmetric  if $l \ne n$ and so the state is  therefore inconsistent with the exclusion principle. The case $\chi=\lambda$ is ignored for having vanishing measure \cite{Jabs10}. Following Feynman \cite{Fey65}, Jabs argues that physical significance resides not in wavefunctions, but rather in transition amplitudes $f$ which must be calculated as
\be\label{Fampl}
f=\int d^4x\int d^4y\,\overline{\Psi_S}\Xi
\ee
where $\Xi=\xi(x,\tau,j,\alpha)\otimes\zeta(y,\tau,k,\beta)$ is a simple tensor product of \wfs for two spin eigenstates respectively having eigenvalues $j, k$ and frame angles $\alpha, \beta$\,. It is readily deduced from (\ref{TWO}) that $f$ is independent of $\tau$\,. In particular the `{\it to}' state $\Psi$ is symmetrized in the sense of (\ref{exd2}), but not the `{\it from}' state $\Xi$\,. If $l=n$ then $f$ is effectively independent of all the frame angles $\alpha, \beta, \chi,\lambda$. The orthogonality of the spin eigenstates implies that, in the case  $l =- n$\,,  the transition amplitude $f$ vanishes if $j=k$.   If $l =-n$ and $j =- k$\,, then $f$ reduces to a single term,  that is, there is no interference between two summands. Furthermore $f$ is then proportional either to unity or to $\kappa$, and so the physically significant $|f|^2$ is independent of the frame angles. The preceding argument would fail if the {\it from} state $\Xi$ were also symmetrized as in (\ref{exd2}), for then the summands in $f$ would not reduce in the case of different spins to a single nonvanishing term which is effectively independent of the frame angles. To continue the construction,  the factor $\kappa$ in (\ref{exd2}) may therefore and without loss of generality be set to unity even if $l \ne n$, yielding the antisymmetric \wf 
\be\label{exd3}
\Psi_{S\,}=\psi (x,\tau, l,\chi)\otimes \phi(y,\tau,n, \lambda)-\phi (x,\tau ,n,\lambda)\otimes \psi(y,\tau ,l, \chi)\,.
\ee
Jabs further points out that, if and only if $\kappa=1$,  the Feynman transition amplitude $f$ in (\ref{Fampl}) coincides with the standard transition amplitude $g$ defined by 
\be\label{Sampl}
g=\int d^4x\int d^4y\,\frac{1}{\sqrt{2}}{\overline{\Psi}}_S\frac{1}{\sqrt{2}}\Xi_S\,.
\ee
Both the {\it from } and {\it to} states are antisymmetrized in $g$, and both are normalized with factors of $1/\sqrt{2}$\,.

\subsubsection{summary of principles}\label{S:prin}
The antisymmetry of the two-particle \wf for indistinguishable \newline spin-1/2 particles, each of which is in either of the two eigenstates of the same spin operator $S(s)$,  follows \cite{Jabs10} from three principles.
\begin{itemize}
\item[P1] The two-particle \wf resulting from an exchange of two particles, both in eigenstates of the same spin operator $S(s)$\,, must be added to the original two-particle \wf for the purpose of calculating transition amplitudes. 

\item[P2] The exchange of  the two unspecified frame-dependent angles must be homotopically consistent, in the case of same spins $(l=n)$ resulting in an antisymmetric two-particle \wf as in (\ref{ex2}) or (\ref{exd3}). In the case of different spins $(l=-n)$ the resulting factor $-\kappa$ as in (\ref{exd2}) may be replaced with negative unity, also yielding an antisymmetric two-particle \wf as in (\ref{exd3})

\item[P3]Transition amplitudes must be calculated as in (\ref{Fampl})  using (\ref{exd3}), that is, after antisymmetrizing the {\it to} state but not the {\it from} state. The result agrees with the standard amplitude  (\ref{Sampl}), for which both two-particle states are exchanged and normalized with factors of $1/\sqrt{2}$\,.

\end{itemize}
Again, the addition principle  in P1 and the transition amplitude in P3 owe to Feynman \cite{Fey65}\,.

\subsection{At $\tau >0$}\label{tau+}

Consider two free and indistinguishable  spin-1/2 particles in plane wave states, both of which are for example positive-energy and  propagating forward in time at $\tau>0$.  Their \wfs  are not necessarily eigenvectors of any spin operators.  It may be assumed that for any one spacelike four-vector $s^\mu$  the two-particle \wf  at parameter $\tau >0$ evolves from a pair of free \wfs for states prepared at  $\tau=0$, both of which \wfs are eigenvectors of the same spin operator $S(s)$. The two-particle \wf at $\tau=0$ must therefore be antisymmetric. The separable free \infl (\ref{Gam02}) preserves antisymmetry, and so the two-particle \wf must be antisymmetric for any $\tau>0$\,. The two-particle \infl is also  \cite{Benn2014} separable in the presence of an electromagnetic field $A^\mu(x)$, and so antisymmetry is preserved even if there is a field in the space-time neighborhoods  of the particles at $\tau>0$\,. If the particles are not in free states at  $\tau >0$, then provided they evolve from states which are free at some $\rho>0$ where $\tau>\rho > 0$, it may be concluded that the two-particle \wf is antisymmetric or fermionic at  $\tau$\,.

\subsection{Other Spins}\label{S:other}
Jabs' argument \cite{Jabs10}  as applied here at $\tau=0$ clearly extends to all half-odd-integer spins, and to all integer spins in which case (\ref{exd3}) takes the symmetric or bosonic form. No fundamental particles of spin greater than 1/2 have been observed. The Higgs boson is the only fundamental spin-0 particle that has been observed \cite{Cho12,Cho12E}. The Klein-Gordon equation \cite{BjDr64} is the relativistic wave equation for spin-0 particles, with  the Stueckelberg wave equation \cite{Stu41a,Stu41b,Stu42} as its parametrized form.  Principle P1  suffices for spin-0 (with the spin operator being unity), since there are no spin frame angles and the transition amplitudes $f$ and $g$ are the same. 

 \section{Discussion}\label{S:disc}
\subsection{Causality}\label{S:caus}
The standard proof of the spin-statistics connection in QFT requires the causality conditions  that are incorporated into  the commutators and anticommutators for bosonic and fermionic fields respectively \cite{Wei95}. The causality conditions are defined by the light cone in $x$ with apex at $y$\,, where $x$ and $y$ are the two events (the arguments of the two fields in the operator products).  The causality conditions in the commutators and anticommutators are also required for $\mathcal{TPC}$ invariance \cite{Sozzi08}, and  for the covariance of Dyson series \cite{Wei95}.   No causality condition is required here for the proof of the spin-statistic connection. Furthermore, $\mathcal{TPC}$ invariance is an immediate consequence of (\ref{PDIR}), while Dyson series for (\ref{PDIR}) are manifestly  covariant since they are $\tau$-ordered rather than $t$-ordered. 

Applications of the parametrized Dirac equation to interactions involve  semiclassical electromagnetic fields, and the fields are constructed from \, M\o ller currents using the standard Feynman propagator which does impose a causality condition \cite{BjDr64,Benn2014}. The single-fermion \infl  $\Gamma^0_+$ in (\ref{plus})  allows free \wfs only of positive mass\footnote{positive-energy particles propagating forward in time, and negative-energy particles backward in time}  to evolve for $\tau$ increasing , and only those of negative mass\footnote{positive-energy particles propagating backward in time, and negative-energy particles forward in time} for $\tau$ decreasing. The detailed form of $\Gamma^0_+(x-x',\tau-\tau')$ does involve the light cone in $x$ with apex at $x'$\,, although propagation is not sharp, but the two-fermion influence function $\Gamma^0_{++}(x-x',y-y',\tau-\tau')$ is separable in $x$ and $y$\,.

\subsection{Entanglement}\label{S:ent}
The first-quantized formalism considered here  is entirely different from QFT. Entanglement is immediately implied by the simple antisymmetric \wf in (\ref{ex2}), and is conserved with respect to $\tau$ by (\ref{TWO}), but is a more complex concept in a quantum field \cite{Summ11}. Some remarks on the nature of entanglement in the two formalisms are therefore in order.

The  parametrized Dirac formalism is not unitary and so an invariant probability density is not available for the definition of entanglement. The antisymmetric \wf $\Psi_S$ in (\ref{ex2}) expresses the entanglement of two  particles for each $\tau$ in the following sense: the covariant current \cite{Benn2014} associated with  $\Psi_S$ is not the sum of the individual currents for the particles. Of all linear combinations of the simple tensor products in (\ref{ex2}), the antisymmetric $\Psi_S$ yields the maximal nonadditivity. The entanglement in $\Psi_S$ is not restricted by $x-y$\,, but the current nonadditivity at $x$ is negligible if the two single-particle \wfs  have negligible overlap  as in $\int d^4y\,\overline{\psi}\phi \backsim 0$\,.

The entanglement of a quantum field itself is expressed in terms of correlation. For a massive vacuum field, the exponential decay scale for correlations with space-like separation is the Compton wavelength  \cite{Summ11}. The proof \cite{Haag58,Araki62,Fred85} assumes that the field operators at $x$ and $y$ (after propagating to a common coordinate time $t$) commute for space-like $x-y$, that is, a causality condition is assumed. It has  been argued \cite{Rez05} that the decay scale of entanglement for two qubits, which have  space-like separation $L$ and which are interacting with a scalar vacuum during a time $T<<L/c$\,, is  $cT$\,. Entanglement for the two qubits is defined as the nonseparability of a density matrix \cite{Wer89}. For times less than $L/c$, the state of one qubit is independent of the other even though there is a correlation in the field \cite{Sab11}. Tests of these remarkable findings have been proposed using circuit QED \cite{Sab12} and cavity QED \cite{Jon14}. The situation here is analogous. The semiclassical electromagnetic field arising from the entangled covariant current is subject to a causality condition, as a consequence of specifying the Feynman propagator. Thus a classical detector will at first receive only the signal from the closer of two entangled particles if the current nonadditivity is negligible. Spin-entangled spatially separated electron pairs can now be produced by splitting Cooper pairs,  with  confirmation of their entanglement approaching feasibility \cite{Scher14}.

\subsection{Summary}\label{S:summ}
Particles of spin-1/2 prepared at the parameter value $\tau=0$ need not satisfy the standard Dirac equation for a sharp mass, even though the wave packets may be extensive in all four dimensions of space-time at $\tau=0$. The particles may therefore all be prepared in  eigenstates of the spin operator $S(s)$ for a single spin axis $s^\mu$ and so, if the particles are indistinguishable and if exchanges of states include homotopically-consistent exchanges of the spin frame angles, Fermi-Dirac statistics are inferred. Subsequent evolution for $\tau>0$ preserves such statistics, even though each particle may no longer be in a spin eigenstate. The relativistic spin-statistics connection obtains here so naturally because the parametrized Dirac equation is  manifestly  covariant as an evolution equation.

\section*{References}
\bibliography{paraDirac_Bib}

\section*{Appendix: linear algebra}\label{S:App}

Given $w_+={\sf w}_+ {\mathbf a}_+=\Lambda_+ w_+$\,, we seek $w_-$  such that $w_-=\Lambda_- w_-$ and $S(s)w=+(1/2)w$ where $w=w_++w_-$\,. The required $w_-$ must satisfy
\be\label{linalg}
Q(s)\Lambda_-w_-=-Q(s)w_+\,,
\ee
which has a solution if and only if  $\overline{z}Q(s)w_+=0$ for all $z$ such that $\overline{z}Q(s)\Lambda_-=0$\,. It suffices to consider a frame in which $s^\mu=(0,0,0,1)$, and so $Q(s)=\mathrm{ diag}(0,1,1,0)$. Assuming $p\cdot s= p_3 \ne 0$, it follows that $z_2=z_3=0$ while $z_1$ and $z_4$ are arbitrary. Hence $\overline{z}Q(s)=0$\,. The solution of (\ref{linalg}) for $w_-$ is undetermined up to the addition of $\Lambda_+b$ for any $b$, but $w_-=\Lambda_- w_-$ is uniquely determined.

Consider two free particles with timelike energy-momenta $p^\mu$ and $q^\mu$\,, where $p \cdot s =p_3 \ne 0$ but $q \cdot s = q_3 =0$\,. There is a boost $\Omega$ to a new frame where $p'=\Omega p$\, and $q'=\Omega q$\,, with $p_3' \ne 0$ and $q_3' \ne 0$. The common spin axis (0,0,0,1) in the old frame may be replaced with (0,0,0,1) in the new frame. The procedure may be performed any finite number of times for any finite number of particles, with $p_3', q_3',\dots$ remaining bounded away from zero. 

 \end{document}